# On Understanding Statistical Data Analysis in Higher Education

**Vera MONTALBANO**
**Department of Physics, University of Siena**
Siena, Italy

**ABSTRACT**

Data analysis is a powerful tool in all experimental sciences. Statistical methods, such as sampling theory, computer technologies necessary for handling large amounts of data, skill in analysing information contained in different types of graphs are all competences necessary for achieving an in-depth data analysis. In higher education, these topics are usually fragmentized in different courses, the interdisciplinary integration can lack, some caution in the use of these topics can missing or be misunderstood. Students are often obliged to acquire these skills by themselves during the preparation of the final experimental thesis. A proposal for a learning path on nuclear phenomena is presented in order to develop these scientific competences in physics courses. An introduction to radioactivity and nuclear phenomenology is followed by measurements of natural radioactivity. Background and weak sources can be monitored for long time in a physics laboratory. The data are collected and analyzed in a computer lab in order to understand the importance of statistical analysis in a not trivial case.

## 1. INTRODUCTION

Data analysis is an important scientific task which is required whenever information needs to be extracted from raw data. It combines several methodologies from statistics, computer science and experimental sciences in which it is widely utilized and developed. Very sophisticated tools are commonly applied in different disciplines, such as high energy physics, medicine, agriculture, biology, economics, environmental sciences, biophysics, and so on. It is possible to use extremely sophisticated data analytic techniques in an attempt to squeeze more and more information out of a given sample of data, or to achieve even more predictive power. But this should not be done out of context. Data analysis must be directed towards solving the real problem, not towards solving a mathematical abstraction of the problem which ignores the fundamental uncertainties intrinsic to the real problem [1].

Many sorts of dangers have become more important as time has passed and computer software has become more powerful. It is all too easy for researchers, no doubt extremely knowledgeable and skilful within their own discipline, to apply very sophisticated data analytic techniques in which they have no expertise whatsoever. The thrust in the development of data analytic software has, quite properly, been in this direction of making software easier to use. In physics, software is often written and utilized by researchers with data analytic expertise joint with physical specific expertise, but this is not always true, especially in other disciplines, so that users are not protected from rushing in and making silly mistakes [1].

In higher education teaching practice it is possible to notice a fragmentation of learning contents related to these topics into different courses. This can be seen in the separation into individual learning subjects, among which exist rare and weak links and what follows is a further separation to individual lessons. Due to the high fragmentation of knowledge, students experience many difficulties in linking information into an understandable, useful and meaningful tool.

Furthermore, introduction to statistics is performed in courses with no link with experimental data and related troubles, thus statistical tools are applied to data sets generated *ad hoc* or very far from student experience. On the other side, statistical tools are often used in laboratory courses without the intent of improving the understanding of their usage. Undergraduate students usually deal with undestanding data analysis by themselves during the preparation of the final experimental thesis, sometimes missing some important aspect. But if they follow a theorical curriculum, it is possible that they never develop this scientific competence. It is not obvious that this lack of knowledge could be recovered in further courses.

In paragraph 2, an early learning path on related topics is presented. The activities were planned and realized in the last class of high school. Remarks and considerations about that experience convinced me to consider a similar path in higher education. Such a proposal is presented in the next paragraph, in order to develop learning paths on statistical data analysis in physics laboratory realized in introductory physics courses for undergraduate students of disciplines different from physics. Finally, a brief discussion for physics students is performed.

## 2. PREVIOUS EXPERIENCES

In the last years, many Italian Universities are involved in National Plan for Science Degree (i.e. Piano nazionale Lauree Scientifiche or PLS), in order to enhance the interest of high school students towards scientific degrees, in particular in Physics, Mathematics and Chemistry.

The PLS guidelines are the following:

1) orienting to Science Degree by means of training
2) laboratory as a method not as a place
3) student must become the main character of learning
4) joint planning by teachers and university
5) definition and focus on PLS laboratories.

There are several types of PLS laboratories. Laboratories which approach the discipline and develop vocations, self-assessment laboratories for improving the standard required by undergraduate courses and deepening laboratory for motivated and talented students.

All activities described in the following were realized in this context, like a PLS laboratory [2] or a part of it [3].

**An early experience**
Since 2006, the Pigelleto's Summer School of Physics is an important appointment for orienting students toward physics. It is organized as a full immersion school on actual topics in physics or in fields rarely pursued in high school, i.e. quantum mechanics, new materials, energy resources. The students, usually forty, are engaged in many activities in laboratory and forced to become active participants [3]. During the 2009 edition, titled *The Achievements of Modern Physics*, a learning path on radioactivity was proposed by me to small groups of students. After a brief introduction to the nuclear phenomenology, they were involved in measures of radioactivity from a weak source of Uranium in order to characterize the emitted ionizing radiation by using an educational device provided by a school.

A teacher was impressed by student's involvement and suggested of elaborating a learning path on nuclear physics in order to propose this activity in her school.

**A learning path on nuclear phenomena in high school**
The activity was supported by the National Plan for Science Degree and the school obtained financial support from a regional project of Tuscany (Progetto Ponte, i.e. Bridge Project) whose purpose was to promote collaboration between Universities and High Schools for orienting students.
The learning path was planned for the last year of high school (5$^a$ Liceo Tecnologico) after the lessons on electromagnetism and about 16 hours were scheduled for the presented activities.
The path consisted in four different but correlated activities [2]: a brief introduction to nuclear phenomena in classroom, measurement of radioactivity in physics laboratory, an introduction to statistical data analysis and some selected elaboration of experimental data performed in computer lab.

**Introduction to nuclear phenomena:** A brief history of discovery of radioactivity was given. The neutron discovery and atom structure were described with a particular attention in student's understanding on dimensions of atom and its constituents. Nuclear radioactivity phenomenology was presented and more details were given for α, β and γ emission. Isotopes, forces into atoms and nuclei were outlined. Transmutations and law of radioactive decay were established. Natural radioactive chains were showed. A special care was made in explaining binding energy and mass of a bound state, equivalence mass-energy and binding energy per nucleon. Finally, fission and fusion were introduced. Topics related to fission and fusion can be very stimulating for students, specially in our society in these times of discussion about usage of nuclear energy and health security.

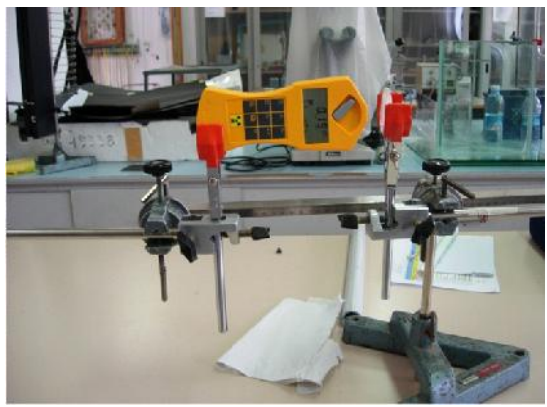

*Figure 1: the experimental set-up*

**Measures of radioactivity:** The measures were realized by means of a commercial Geiger counter usually used for dosimetry. It can separate different contribution from α, β and γ rays and data were collected without any unit conversion and downloaded into a computer. The detector and the sources were aligned on an optical bench as shown in Fig. 1. The sources were small Uranium sheets ($m_1$ = 0.152 g), another Uranium sheets ($m_2$ = 0.888 g) and a very weak Uranium ore source (fluorescent marble). Students realized background measures, measurement of Uranium sources at different distances, α, β and γ rays measurement, the previous ones with different shields, measures with all three sources.

**Introduction to statistical data analysis:** A question was proposed: there are limits to the precision of a measure? By answering to this question, one can fix the cases in which the use of statistical data analysis make sense. The first step was to recall all previous knowledge of students on uncertainties and statistics [4] , which are many but almost never used in physics by teachers. Starting from the beginning, main topics were remembered with examples, such as representation of a casual phenomena, mean, median, variance and standard deviation, Bernoulli, Poisson and Gauss distributions At this point, it was possible to introduce some elements of sampling theory [5], such as existence of a parent distribution, characterization of a good statistical sample, its mean and variance. Many links to experimental data collected in physics laboratory were proposed and examples were constructed by using them, when it is possible.

**Elaboration in computer lab:** Many elaborations were proposed in order to improve student's understanding of statistical data analysis. Let me give some example: verify that radioactive decay follows Poisson's distributions, test different samples by using 10, 20, 30, 50 data from the background, compare means and standard deviations of these samples with the mean and standard deviation of very long measure, find for which values of mean the Poisson's distribution becomes indistinguishable from the normal one. All possible informations are extracted from experimental data. Finally, slightly different measures were compared. The raw data are showed in Fig, 2.

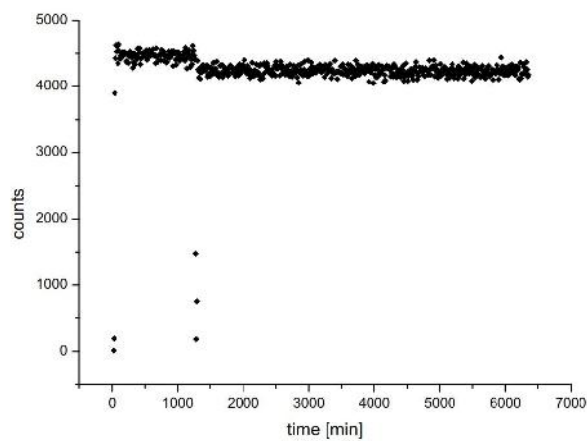

*Figure 2: measures of ionizing radiation emitted from a source of Uranium at two different distances from the Geiger detector, counts are summed and registered every ten minutes*

It is possible to recognize two different set of data ( $N_1$ = 121, mean 4472, standard deviation 71, sample standard deviation 6 and $N_2$ = 506, with mean 4238, standard deviation 67, sample standard deviation 3). It is easy to compute the difference between means but it is overwhelmed by the uncertainty if one uses standard deviation, on the contrary the use of sample standard deviation allows to estimate quantitatively the difference $N_1 - N_2 = 226 \pm 18$.

**Remarks:** The learning path described above was performed on three last classes in 2010 and 2011. All classes made the introduction to nuclear phenomena in class and in laboratory, only one class made some activity in computer lab. In my opinion, this path is very interesting for students but hardly feasible in last class due to graduation exam. It can be an excellent optional proposal for interested students but if a teacher want to implement it in his classes, the presentation of statistical topics in previous years needs to be revised.

### 3. SOME PROPOSAL IN HIGHER EDUCATION

The above considerations led me to think about the possibility of proposing similar paths for undergraduate students. Nuclear phenomena remains one of more exciting topics rarely treated in introductory courses of physics. Some informal interview with our PhD students and some physics faculty convinced me that almost nothing is changed in Italian higher education in teaching this interdisciplinary topic, despite all curricula are revised many times in the last ten years.

**A learning path on nuclear phenomena in introductory physics courses**
In many undergraduate courses of study, such as Medicine, Biology, Environmental or Geological Sciences, physics remains a basic issue and one or more introductory physics courses are compulsory in the curricula.

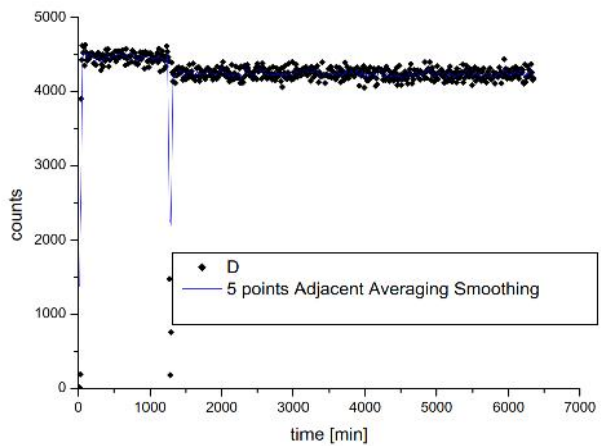

*Figure 3: raw data and smoothing curve*

Nevertheless, in these courses there is not enough time for offering a complete and earnest learning path on classical physics. Furthermore, vary other topics are requested such as some elements of applied or modern physics, often with the possibility of using physics laboratory. Statistics is usually taught by a mathematician and no link to experimental data is made.

Another problem that can arise is teaching in very large classes (greater then 100). In this situation, it is not easy to choose which topics are more effective and useful for these students. In my opinion, it could be a good choice to dedicate part of a physics course to a learning path similar to that described in the previous paragraph. Nuclear phenomena can be chosen as experimental topic, but other choices may be equally valid, or perhaps more suited to certain curricula. For instance, for student in environmental sciences a noise pollution measurement or nuclear measures on radon pollution can be more interesting. What matters is that experimental data must be referred to measures of a meaningful physical quantity.

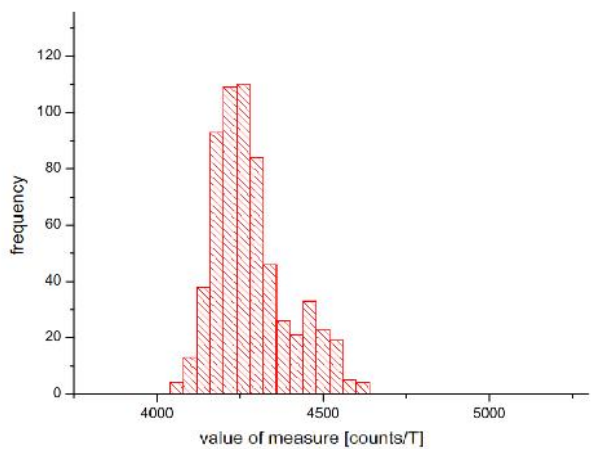

*Figure 4: histogram of frequencies (T = 10 min)*

Large classes can make this activity too, because it is possible to show the functioning of measuring devices to all students which can perform measures divided in small groups at different times. Since these students have more mathematical and scientific competences compared to students in high school, one can utilizes more advanced instrumentation for measuring physic al quantities, collecting and elaborating experimental data. For instance, the data showed in Fig.2, can be analyzed by using a common software for fitting data (e. g. Origin) and more sophisticated statistical tools can be used. In Fig. 3, a smoothing of data is performed. The histogram of frequencies of total data is showed in Fig. 4. It is easy to recognize that it is not a good statistical sample. Nevertheless, if a fit with two normal curves is performed (Fig. 5), one can understand that it is the convolution of two normal distributions.

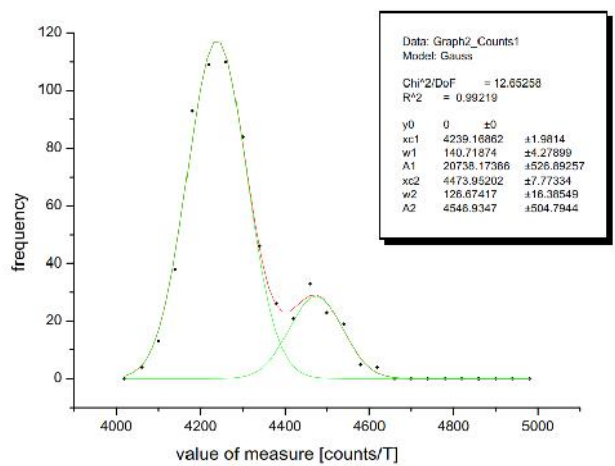

*Figure 5: fit of histogram of frequencies with two normal curves*

Thus, data are obviously the union of two good statistical samples with different means and standard deviations which are different from that obtained at the end of previous paragraph. Discussions with students about the origin of this behavior can be very instructive for a depth understanding of statistical data analysis.

**A different learning path on data analysis for physics laboratory in course of study in basic sciences**

In the case of undergraduate courses of study for basic sciences, such as Physics, Astronomy or Chemistry, the situation is more complex. Physics laboratories are very important courses, together with introductory, and sometimes advanced, courses in statistical data analysis. Different curricula can prepare in a very different way students. Usage of statistical tools can be really advanced for researchers in some discipline, like for High Energy Physics. An instance in this case is shown in Fig. 6 [6].

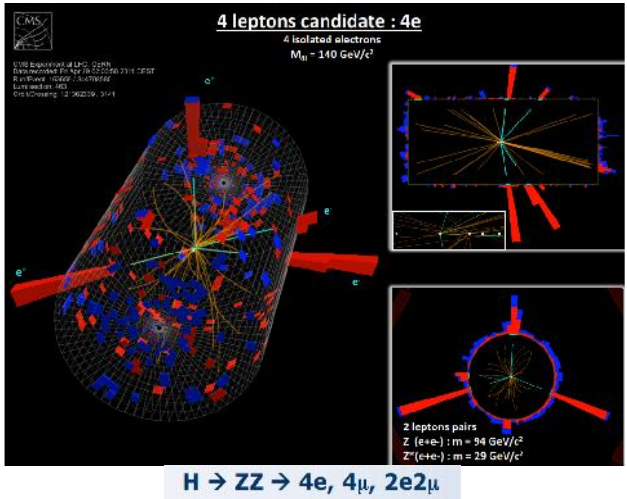

*Figure 6: a four leptons candidate for Higgs Boson decay from Compact Muon Solenoid Experiment at Large Hadron Collider, CERN*

In Fig. 6, different graphs from statistical data analysis of the event merged with pictorial schemes of the detectors set-up are shown .

How many data have been collected and analysed for obtaining a candidate on this hot research topic? The obvious answer is a huge amount. A raw estimate of this magnitude is given by integrated luminosity of LHC at the energy of 7 TeV times the inelastic cross section for proton-proton collision measured from TOTEM Experiment (Total Cross Section, elastic scattering and diffraction dissociation measurement) at LHC. The result is that, in order to obtaining about 25 candidates for Higgs boson decay in four leptons, as the one showed in figure, it was necessary to produce and consider $350 \times 10^{12}$ events, i.e. less than one useful event every 10000 billions of produced events [7].

The researchers in this discipline must have, of course, a real solid formation in data analysis that usually develop by themselves in large international teams. It is singular that their training in this field is fragmented during their course of study and almost completely postponed to further research experience. Work is in progress with the purpose of conceiving a meaningful learning path on data analysis both in undergraduate and graduate courses for these kind of students.

## 4. CONCLUSIONS

Some proposal in higher education for realizing a learning path on statistical data analysis strongly linked to physics laboratory has been displayed. The next step will be to test at least one of these learning paths in undergraduate courses in order to proof their effectiveness. More care must be posed in elaborating new learning path in data analysis for basic sciences. The case of physics was outlined in previous paragraph but they exist other delicate and peculiar contexts, such as Mathematics or Pharmacy, that remain unexplored.

## 5. ACKNOLEDGEMENTS

This work is based on previous activities and experiences which were performed within the National Plan for Science Degree supported by Italian Ministry of Education, University and Research. The author would like to thank Angelo Scribano for the active support and useful discussions.